\documentclass[aps,prb,floatfix,amsmath,amssymb,preprint,superscriptaddress]{revtex4}
\usepackage[dvips]{graphicx}
\usepackage{dcolumn}
\usepackage{bm}
\begin{document}
\title{Fermi surface change across a\\ deconfined quantum critical point}

\author{Ribhu K. Kaul}
\affiliation{Department of Physics, Harvard University, Cambridge MA
02138}

\author{Alexei Kolezhuk}
\affiliation{Department of Physics, Harvard University, Cambridge MA
02138}
\affiliation{Institut f\"ur Theoretische Physik,
Universit\"at Hannover, 30167 Hannover, Germany}

\author{Michael Levin}
\affiliation{Department of Physics, Harvard University, Cambridge MA
02138}

\author{Subir Sachdev}
\affiliation{Department of Physics, Harvard University, Cambridge MA
02138}

\author{T. Senthil}
\affiliation{Center for Condensed Matter Theory, Department of
Physics, Indian Institute of Science, Bangalore 560 012, India}
\affiliation{Department of Physics, Massachusetts Institute of
Technology, Cambridge MA 02139}

\begin{abstract}
The quantum phase transitions of metals have been extensively
studied in the rare-earth ``heavy electron'' materials, the
cuprates, and related compounds \cite{coleman0}. The Fermi surface
of the metal often has different shapes in the states well away from
the critical point. It has been proposed
\cite{coleman0,qsi0,senthil0} that these differences can persist up
to the critical point, setting up a discontinuous Fermi surface
change across a continuous quantum transition. We study square
lattice antiferromagnets undergoing a continuous transition from a
N\'eel state to a valence bond solid, and examine the fate of a
small density of holes in the two phases. Fermi surfaces of charge
$e$, spin $1/2$ quasiparticles appear in both phases, enclosing the
usual Luttinger area. However, additional quantum numbers cause the
area enclosed by each Fermi surface pocket to jump by a factor of 2
across the transition in the limit of small density. We demonstrate
that the electronic spectrum across this transition is described by
a critical theory of a localized impurity coupled to a 2+1
dimensional conformal field theory. This critical theory also
controls the more complex Fermi surface crossover at fixed density,
which likely involves intermediate phases with exotic fractionalized
excitations. We suggest that such theories control the electronic
spectrum in the pseudogap phase of the cuprates \cite{mohit}.
\end{abstract}




%

\date{\today}

\maketitle

\section{Introduction}

Many strong correlation problems of current interest relate to the
physics of doped Mott insulators. Most observed metallic states can
be understood in conventional terms: ordering associated with broken
spin or lattice symmetries, along with Fermi surfaces of charge $e$,
spin 1/2 quasiparticles. Nevertheless, the details of the state
differ depending upon the importance we attach to strong correlation
physics {\em e.g.\/} different Fermi surface topologies and magnetic
moment configurations are obtained from the weak and strong
interaction limits, namely ({\em i\/}) a spin density wave
instability of the free electron band structure, or ({\em ii\/}) a
spin-wave theory of a lightly doped magnetically ordered Mott
insulator. It is clear that rich physics lies in understanding the
interpolation between such limiting regimes, and that there a number
of important applications {\em e.g.\/} to the widely discussed
differences between the underdoped and overdoped cuprates.

A central question is whether there could be a single continuous
quantum phase transition between such states with distinct types of
order or Fermi surface topologies (recent experiments \cite{onuki}
on CeRhIn$_5$ are compatible with an abrupt or very rapid change in
Fermi surface topology). Such critical points could be the key to
understanding the many observed `non-Fermi liquid' properties.
However, a complete critical field theory of a continuous transition
between such states has so far been presented only in
insulators~\cite{senthil1}, where the two states have broken
symmetries which cannot be connected by a continuous transition in
Landau theory; instead a `deconfined' theory focussing not on order
parameters but on fractionalized excitations and emergent gauge
forces shows that such transitions are indeed possible. An extension
of these ideas to metallic systems has been discussed
\cite{ffl,senthil0}, including a scenario which would allow for a
discontinuous change in a Fermi surface. Others \cite{qsi1} have
proposed a `local quantum critical' scenario for transitions between
states with different Fermi surface topologies, based upon numerical
studies in extensions to dynamic mean field theory. However, the
analytic structure of this mean field transition remains unclear,
and it is not clear how, even in principle, this `local critical'
structure can be extended beyond mean field theory.

\begin{figure}[p]
\centering
  \includegraphics[width=6.5in]{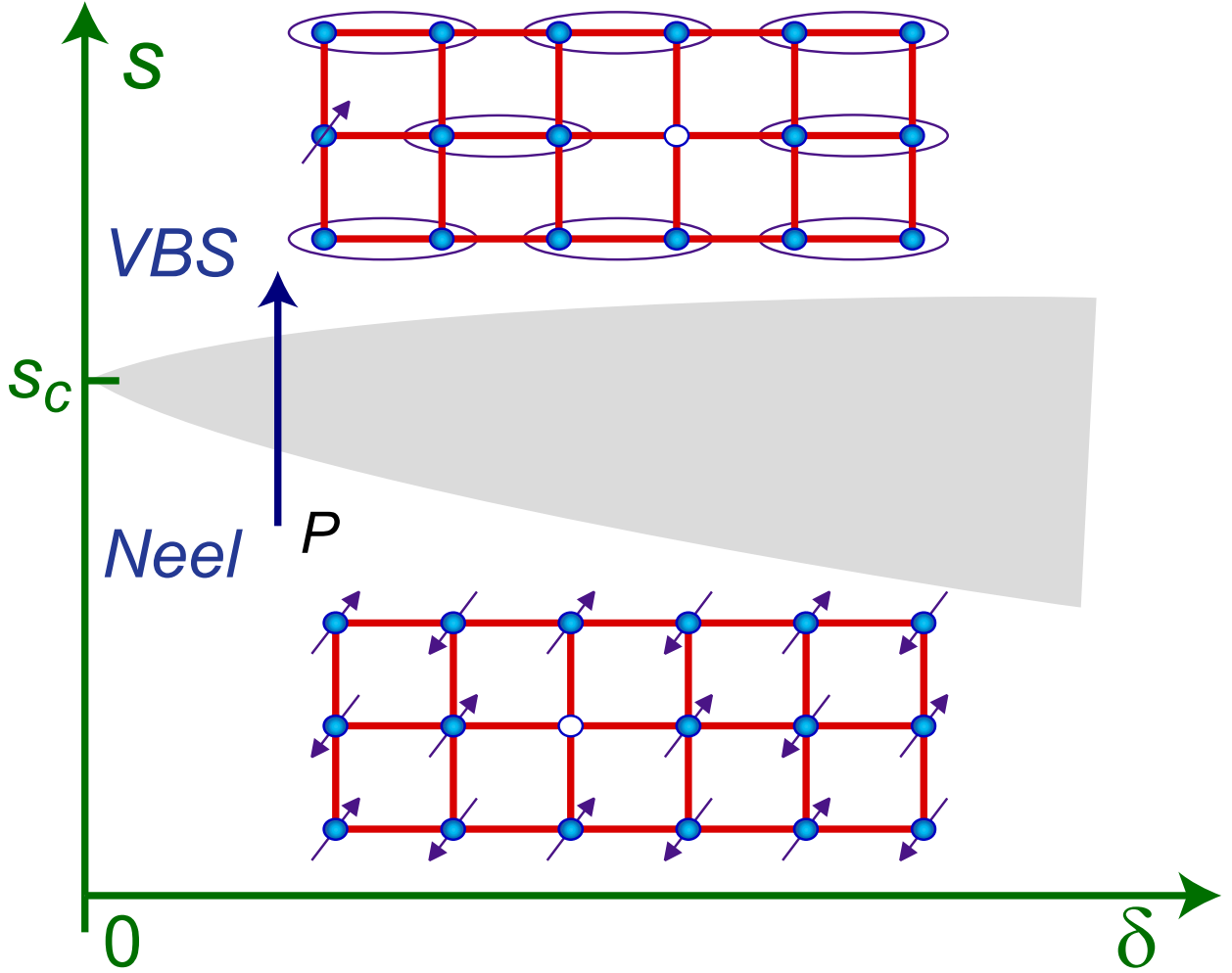}
  \caption{Schematic phase diagram. The ellipses represent spin singlet valence
  bonds. The coupling $s$ tunes the insulator across the N\'eel-VBS
  transition, and $\delta$ is the mobile hole density.
  The deconfined quantum critical point is at $s=s_c$ in the
  insulator with $\delta=0$. The vacancies (`holons') carry a gauge charge $q=\pm 1$
  under an emergent U(1) gauge force. In the cartoons above,
  the reader can interpret $q$ as a sublattice label.
  In the $s < s_c$, N\'eel phase,
  $q$ determines the spin: a vacancy on an up (down) spin site
  carries net spin down (up) and so is equivalent to a charge $e$ spin-1/2 hole.
  For $s>s_c$, the hole is a composite of a vacancy and
  a nearby unpaired spin with opposite $q$, moving by rearranging
  nearest-neighbor valence bonds; note that this motion preserves spin and sublattice quantum
  numbers {\em separately\/} (see also Ref.~\onlinecite{ls}). So
  there are twice
  as many states per momentum for a charge $e$ spin 1/2 hole in the VBS state than there are in the N\'eel state.}
  \label{fig1}
\end{figure}
\begin{figure}[p]
\centering
  \includegraphics[width=6.5in]{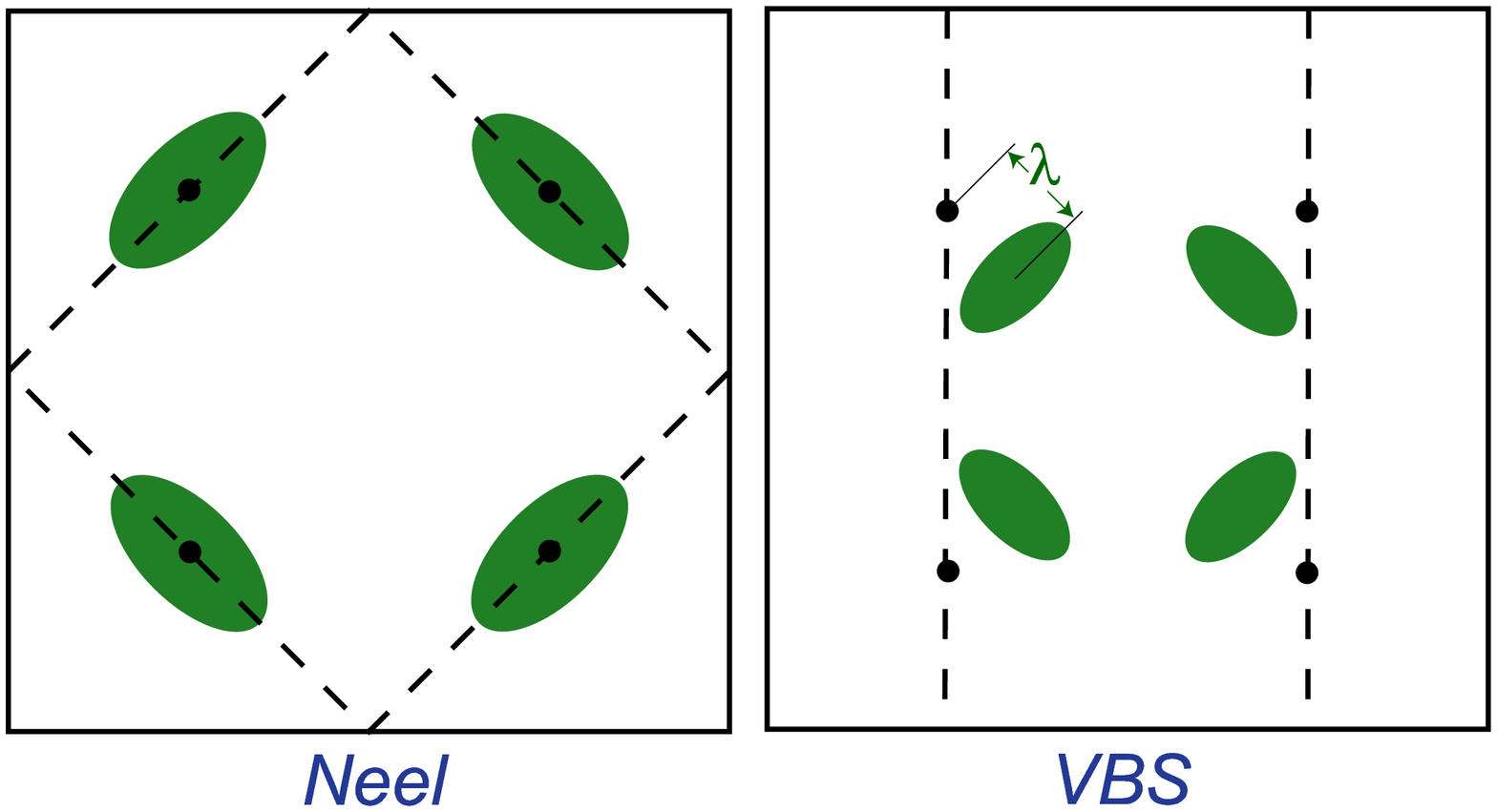}
  \caption{Momentum space Fermi surfaces in the N\'eel and VBS regions of
  Fig~\ref{fig1}. The filled circles are the 4 $\vec{K}_p$
  wavevectors, with $\vec{K}_1 = (\pi/2a) (1,1)$, $\vec{K}_2 =
  (\pi/2a) (1,-1)$, $\vec{K}_3 = - \vec{K}_1$, $\vec{K}_4 = -
  \vec{K}_2$, with $a$ the lattice spacing.
  The dashed line in the N\'eel phase indicates the boundary of the
  magnetic Brillouin zone. Only the Fermi surfaces within this zone
  contribute to the Luttinger counting, and so the area of each
  ellipse is $\mathcal{A}_F = (2 \pi)^2 \delta/4$. In the VBS phase,
  all 4 pockets are inequivalent, and so the area of each ellipse is
  $\mathcal{A}_F = (2 \pi)^2 \delta/8$. The dashed lines now show
  the reduction of the Brillouin zone due to the VBS order which appears at sufficiently
  low temperatures;
  ``shadow'' Fermi sufaces, with weak photoemission intensity
  (estimated in the text), will appear as
  reflections across these lines, and these Fermi surfaces are not
  shown.
  }
  \label{fig2}
\end{figure}
In this paper, we will build on the theory of the deconfined quantum
critical point proposed in Ref.~\onlinecite{senthil1} for insulating
spin $S=1/2$ antiferromagnets on the square lattice
\cite{proko,anders,kamal,mv}; the transition is between (see
Fig.~\ref{fig1}) a state with magnetic N\'eel order (with spin
polarized in the staggered configuration) to a spin-gap state with
valence bond solid (VBS) order (which is spin rotation invariant,
but breaks lattice symmetries by ordering of valence bonds). We will
describe the metallic states obtained by doping this system with a
small density ($\delta$) of holes. Figs.~\ref{fig1} and~\ref{fig2}
summarize our key results, and the captions support them with simple
physical arguments. The decoupling of the spin and U(1) gauge charge
in the VBS phase described in Fig.~\ref{fig1}, and the consequent
doubling of the hole flavors in Fig.~\ref{fig2}, leads to our
precise result: the discontinuity in $\lim_{\delta \rightarrow 0}
\mathcal{A}_F/\delta$ on the two sides of the N\'eel-VBS quantum
critical point, as shown in Fig.~\ref{fig2}.

On the other hand, the evolution of the Fermi surface along the
fixed $\delta$ line $P$ in Fig~\ref{fig1} involves separate
questions associated with the interplay between the mobile holes and
the $s=s_c$ deconfined critical point. If the $\delta=0$
monopole-induced confinement physics \cite{senthil1} survives for
$\delta > 0$, then there will be a single critical line which is
crossed by $P$ at which the Fermi surface changes discontinuously.
However, this appears unlikely because the monopoles can be screened
by the charge carriers \cite{hermele}, leading to a narrow
intermediate fractionalized `holon metal' phase with no conventional
Fermi surface within the shaded region. In either case, there is a
wide momentum, frequency, and temperature regime over which the hole
spectral function is incoherent. We will show here that this is
described by a universal theory of a single localized vacancy
\cite{qimp} coupled to the 2+1 dimensional conformal field theory of
the $\delta=0$, $s=s_c$ critical point. The incoherent, critical
hole spectral function is momentum-independent, and so while this
view of the transition has a `local' character \cite{qsi1}, the
degrees of freedom include a non-trivial, bulk quantum field theory.
Further, we believe that the evolution of the fermionic excitations
across the shaded region of Fig~\ref{fig1} cannot be captured by
dynamic mean field theory methods or self-consistent impurity
models.

\section{Field theory}

The motion of charge carriers doped into an insulating
square-lattice quantum antiferromagnet is conventionally described
by the ``$t$-$J$'' model,
\begin{equation}
\label{tJmodel}
H_{t-J} =
- \sum_{i,j,\sigma}t_{ij}(c^\dagger_{i\sigma}c_{j\sigma} + {\rm h.c.})
+  \sum_{i,j} J_{ij}\vec{S}_i \cdot \vec{S}_j,
\end{equation}
where $c^\dagger_{i\sigma}$ creates an electron with spin $\sigma$
on the sites $i$ of a square lattice and
$\vec{S}_i=\frac{1}{2}\sum_{\sigma\sigma^\prime} c^\dagger_{i\sigma}
\vec{\sigma}_{\sigma\sigma^\prime}c_{i\sigma^\prime}$. In addition,
the constraint $\sum_{\sigma}c^\dagger_{i\sigma}c_{i\sigma}\leq 1$
is enforced on each site, modelling the large local repulsion
between the electrons. It is important to note that our results are
more general than a particular $t$-$J$ model, and follow almost
completely from symmetry considerations.

We begin our analysis by recalling the physics in the absence of
doped holes; we then only need to consider the second term in
Eq.~(\ref{tJmodel}) {\em i.e.\/} the ``Heisenberg model''. With
weakly frustrated $J_{ij}$ the ground state is the N\'eel state.
With stronger frustration (or additional ring exchanges), there can
be a quantum transition to a VBS phase\cite{rs}. The field theory
for the N\'eel-VBS transition in the insulator \cite{senthil1} is
expressed in terms of a complex `relativistic' field $z_\alpha$,
where $\alpha=\uparrow,\downarrow$, which is the field operator for
a charge neutral, spin $S=1/2$ spinon excitation. Low lying singlet
excitations are represented by an emergent U(1) gauge field $A_\mu$,
where $\mu=\tau,x,y$ is a spacetime index. The field theory has the
action $\int d^2 r d\tau \mathcal{L}_z$ with
\begin{equation}
\mathcal{L}_z = |( \partial_\mu - i A_\mu ) z_\alpha |^2 + s
|z_\alpha |^2 + \frac{u}{2} \left( |z_\alpha|^2 \right)^2 + \ldots
\label{lz}
\end{equation}
Here $s$ is the tuning parameter in Fig.~\ref{fig1}, we have set a
spinon velocity to unity, and $u$ is a spinon self-interaction. For
$s>s_c$, this field theory is in a spin SU(2)-invariant U(1) spin
liquid phase, with gapped spinon excitations, and provides a
continuum description of the Schwinger boson mean field theory of
the square lattice antiferromagnet \cite{aa}. The inclusion of
monopole tunneling events shows that this spin liquid is ultimately
unstable to confinement and the development of VBS order \cite{rs}.
This monopole physics is peripheral to our main results, and so we
defer its discussion until later.

Now we need to add charge carriers to this antiferromagnet.
We have represented the spin excitations of the antiferromagnet by
Schwinger bosons above, and so the charge carriers require a
spinless fermion $f$, the `holon' (see Appendix~\ref{app:psg} for
more details). Such a mean-field lattice study has been carried out
some time ago in Refs.~\onlinecite{klr,suyu}. We can take the low
energy limit of their results, and so obtain the field theory needed
for our analysis. However, we can also proceed on symmetry grounds,
when our main ingredients are the transformation properties of the
various fields under square lattice symmetry operations. Because the
elementary quanta also carry U(1) gauge charges, it is not required
that the fields be invariant under such operations, only their
gauge-invariant combinations. This larger space of transformations
was dubbed the projective symmetry group (PSG) by Wen \cite{wenpsg},
and we list the PSG of the $z_\alpha$ in Table~\ref{tablepsg}. This
$f$ holon also carries the gauge charge of the U(1) gauge field
$A_\mu$ in Eq.~(\ref{lz}). Indeed, because the spinon field flips
charges under a sublattice interchange in Table~\ref{tablepsg}, we
see that holons on the two sublattices have charges $q=\pm 1$. This
$q$ quantum number will play a central role. However, the holons can
also acquire an additional `valley' quantum number if their
dispersion minimum is not at the center of the Brillouin zone; we
will keep track of this at the cost of some additional bookkeeping.
The analysis of the holon Green's function of the present U(1) spin
liquid phase, along the lines of Ref.~\onlinecite{suyu}, shows that
there are 2 distinct valleys, with labels $v=1,2$, and we choose a
gauge in which their minima are at wavevectors $\vec{K}_{1,2}$ shown
in Fig.~\ref{fig2}. Summarizing, we have 4 species of spinless
fermions $f_{qv}$, with U(1) charge $q=\pm 1$, and valleys $v=1,2$.
All the information we need to construct the effective action and
the observable correlations of these fermionic holons is
encapsulated in their PSG, which is derived in
Appendix~\ref{app:psg}, and is summarized in Table~\ref{tablepsg}.
\begin{table}[t]
~~\\~~\\ \centering
\begin{tabular}{||c||c|c|c|c||} \hline\hline
 & $T_x$ & $R_{\pi/2}^{\rm dual}$ & $I_x^{\rm dual}$ & $\mathcal{T}$  \\
 \hline\hline
$z_\alpha$ & $ \varepsilon_{\alpha\beta} z^{\beta \ast}$  & $
\varepsilon_{\alpha\beta} z^{\beta \ast}$ & $
\varepsilon_{\alpha\beta} z^{\beta \ast}$ & $
\varepsilon_{\alpha\beta} z^{\beta }$
\\
\hline $f_{+1}$ & $i f_{-1}$ & $i f_{-2}$ & $i f_{-2}$  &  $f_{+1}$ \\
\hline $f_{-1}$ & $-i f_{+1}$ & $i f_{+2}$ & $i f_{+2}$  & $ -f_{-1}$ \\
\hline $f_{+2}$ & $i f_{-2}$ & $i f_{-1}$ & $i f_{-1}$  &  $f_{+2}$ \\
\hline $f_{-2}$ & $-i f_{+2}$ & $-i f_{+1}$ & $i f_{+1}$  &  $- f_{-2}$ \\
\hline $\Phi_{\rm VBS}$ & $-\Phi_{\rm VBS}^\ast $ & $i\Phi_{\rm
VBS}^\ast$ &
$\Phi_{\rm VBS}$ & $\Phi_{\rm VBS}$ \\
\hline\hline
\end{tabular}
\caption{PSG transformations of the fields under square lattice
symmetry operations. $T_x$: translation by one lattice spacing along
the $x$ direction; $R_{\pi/2}^{\rm dual}$: 90$^\circ$ rotation about
a dual lattice site on the plaquette center ($x\rightarrow
y,y\rightarrow-x$); $I_x^{\rm dual}$: reflection about the dual
lattice $y$ axis ($x\rightarrow -x,y\rightarrow y$); $\mathcal{T}$:
time-reversal, defined in real time. $\varepsilon_{\alpha\beta}$ is
the antisymmetric tensor with $\varepsilon_{\uparrow\downarrow}=1$
and $\varepsilon_{\alpha\beta}=\varepsilon^{\alpha\beta}$. The local
VBS order parameter is defined as $\Phi_{\rm VBS}(\vec{r})\equiv
(-1)^{\vec{r}\cdot \hat{x}}\langle S_{\vec{r}} \cdot S_{\vec{r}+
\hat{x}}\rangle + i (-1)^{\vec{r}\cdot \hat{y}}\langle S_{\vec{r}}
\cdot S_{\vec{r}+\hat{y}}\rangle$; where $\hat{x},\hat{y}$ are the
unit lattice vectors. } \label{tablepsg}
\end{table}

An immediate consequence of the PSG of the $z_\alpha$ and the
$f_{qv}$ is that we can write down an expression for the physical
charge $-e$, spin-1/2 electron annihilation operator $c_{\alpha}
(\vec{r}) = \sum_{p=1}^4 e^{i \vec{K}_p \cdot \vec{r}}
\Psi_{p\alpha} (\vec{r})$ where the $\vec{K}_p$ are defined in
Fig.~\ref{fig2} and
\begin{eqnarray}
\Psi_{1,3\alpha} &=& z_\alpha f_{+1}^\dagger \pm
\varepsilon_{\alpha\beta}
z^{\beta \ast} f_{-1}^\dagger \nonumber \\
\Psi_{2,4\alpha} &=& z_\alpha f_{+2}^\dagger \pm
\varepsilon_{\alpha\beta} z^{\beta \ast} f_{-2}^\dagger \label{psi}
\end{eqnarray}
These relations will allow us to construct the Fermi surfaces in
Fig.~\ref{fig2}.

The field theory of the doped antiferromagnet has the Lagrangian is
$\mathcal{L} = \mathcal{L}_z + \mathcal{L}_f + \mathcal{L}_c$. The
fermionic holon terms are
\begin{equation}
\label{eq:Lf}
\mathcal{L}_f =  \sum_{qv} f^\dagger_{qv}  \biggl( \partial_\tau - i
q A_\tau -\mu - \frac{(\partial_{\overline{j}} - i q A_{\overline{j}} )^2}{2m_{v\overline{j}}} \biggr)
f_{qv},
\end{equation}
where $\overline{j}$ extends over $\overline{x},\overline{y}$,
$m_{1\overline{x}}=m_{2\overline{y}}$ and
$m_{2\overline{x}}=m_{1\overline{y}}$ are the mass of the elliptical
hole pockets, $\mu$ is the hole chemical potential, and the
$\overline{x}$ and $\overline{y}$ directions are rotated by
45$^\circ$ from the principle square axes. The PSG prohibits a
linear derivative term in Eq.~(\ref{eq:Lf}), and so the $f_{qv}$
dispersion is an extremum at zero momentum; this pins the Fermi
surfaces to be centered at $\vec{K}_p$ in the N\'eel phase
(Fig.~\ref{fig2}), but not, as we will see, in the VBS phase.
Finally, we include coupling between the holons and spinons
invariant under all PSG operations. There is an unimportant scalar
coupling $\sum_{\alpha q v} |z_\alpha|^2 f^\dagger_{qv} f_{qv}$, but
the more important term is
\begin{equation}
\mathcal{L}_c = i \widetilde{\lambda} \varepsilon^{\alpha\beta}
\left\{ f^\dagger_{+1} f_{-1} z_\alpha
\partial_{\overline{x}} z_\beta  + f^\dagger_{+2} f_{-2} z_\alpha
\partial_{\overline{y}}  z_\beta  \right\} + \mbox{c.c.},
\label{lc1}
\end{equation}
equivalent to the dipole coupling introduced by Shraiman and Siggia
\cite{ss}, arising from the hopping of electrons between
nearest-neighbor sites. Another symmetry-based derivation of the
theory of the doped antiferromagnet was recently given \cite{wiese}
for the N\'eel phase alone. In contrast, our analysis starting from
a spin rotation invariant spin liquid can address the quantum
critical point, and can also be specialized to the N\'eel state by
spontaneously breaking spin rotation symmetry. We now turn to an
analysis of the motion of holes in the field theory
$\mathcal{L}=\mathcal{L}_z + \mathcal{L}_f + \mathcal{L}_c$, first
in the N\'eel and VBS phases, and then finally at the deconfined
quantum critical point that seperates these two phases.

\section{Doping the N\'eel state}

First, consider the $s<s_c$ N\'eel phase of $\mathcal{L}_z$ with
$\langle z_\alpha \rangle \neq 0$. By the Higgs mechanism, the
condensate of $z_\alpha$ renders the $A_\mu$ massive, and so we can
safely integrate the $A_\mu$ out. A crucial point is that the Higgs
mechanism also ties the U(1) gauge charge $q$ of the holons to the
spin quantum number $S_z$ along the N\'eel order (because of the
broken symmetry, $S_x$ and $S_y$ are not conserved); more precisely
$S_z = q/2$, and so the $f_{qv}$ carry all the quantum numbers of
the electron. This is established in Appendix~\ref{app:neel}.

Now we consider the Fermi liquid state obtained with a total density
$\delta$ of the four fermion species $f_{qv}$. Each will form a
separate Fermi surface containing $\delta/4$ states; their coupling
to the spin-wave can be treated perturbatively, and do not modify
any of the key Fermi liquid characteristics. From Eq.~(\ref{psi}) we
deduce that there are 4 hole pockets centered at the $\vec{K}_p$
wavevectors, each enclosing the area $\mathcal{A}_F = (2\pi)^2
(\delta/4)$, as shown in Fig.~\ref{fig2}. The caption of
Fig.~\ref{fig2} shows that the same area is obtained in direct Fermi
liquid counting of electrons within the magnetic Brillouin zone.

\section{Doping the VBS state}

Next, we examine doping the $s>s_c$ state. For now, we neglect
monopoles, and so the undoped state is a spin rotation invariant
U(1) spin liquid with $A_\mu$ representing a gapless `photon'
excitation; the spinful excitations are gapped, but single $S=1/2$
spinons do not exist as asymptotic states---the logarithmic
`electrostatic' potential mediated by the $A_\tau$ binds the spinons
in pairs. We can identify $2 \Delta$ as the gap towards creating the
lowest spinful excitation; this vanishes at the quantum critical
point as $\Delta \sim |s-s_c |^\nu$. In a theory with $N$ spinon
species ($\alpha = 1 \ldots N$ in Eq.~(\ref{lz})), the logarithmic
potential will have the form $V(r) = (12 \Delta/N) \ln (r \Delta)$
in the large $N$ limit \cite{gm}. Now add a single holon into this
state. The coupling to $A_\tau$ causes this hole to have an
`electrostatic' self energy which diverges logarithmically with
system size \cite{qimp}, and so a spin-singlet single holon state is
not stable. Rather, the holon will peel off a single spinon from
above spinon gap, and form a $S=1/2$, charge $e$ bound state
\cite{qimp}. This bound state has the same quantum numbers of as an
ordinary hole, and is neutral under the $A_\mu$ U(1) gauge force. A
finite density of such bound states can then form a Fermi surface
with charge $e$ $S=1/2$ quasiparticles. We also have to consider the
pairing between holons with opposite U(1) gauge charges, induced by
the $A_\mu$ gauge force (this attractive force must be balanced
against the repulsive Coulomb force associated with the physical
electromagnetic charge $e$ of each holon). This is expected to lead
to superconductivity at low enough temperatures, but we will not
discuss this here; our focus is on the normal state.

The structure of the holon-spinon bound state is described in more
detail in Appendix~\ref{app:bound}. As a consequence of the mixing
between the different bound states induced by the terms in
$\mathcal{L}_c$, the bound state eigenmodes turn out to be precisely
the electron-like momentum eigenstates in Eq.~(\ref{psi}), with a
dispersion which has a minimum at $K_{\overline{x}} = \pm \lambda$ where $\lambda
 \sim \widetilde{\lambda} \Delta^2$. This is responsible for the
shift in the centers of the elliptical hole pockets shown in
Fig.~\ref{fig2}.

The usual counting argument now allows us to deduce the area of each
hole pocket. There are 4 inequivalent pockets, and a factor of 2
degeneracy for spin; so $\mathcal{A}_F = (2 \pi)^2 (\delta /8)$. The
factor of 2 difference from the result obtained in the N\'eel phase
is one of our key results.

The state we have described so far is actually {\em not\/} a
conventional Fermi liquid: the area enclosed by its Fermi
surface is not the same as that of a non-interacting electron gas
at the same filling and with the same size of unit cell. In the
non-interacting case one would have a Fermi surface area
$(2\pi)^2(1-\delta)/2$.
  In our case, where the holes were doped on a background spin
liquid state with a gapless photon excitation, we obtain that the
electron-like area enclosed by the Fermi surface is $(2\pi)^2 (1-\delta/2)$
. This state is a {\em fractionalized Fermi liquid\/}
\cite{ffl}, obtained here in a single band model, in contrast to its
previous appearance in Kondo lattice models.

The instability of the spin liquid to confinement and VBS order
induced by monopoles \cite{rs,senthil1}, and the associated halving
of the Brillouin zone will finally transform the state into one
obeying the conventional Luttinger theorem, with an electron-like
area enclosed by the Fermi surface of $(2\pi)^2 (1 - \delta)/2$. The
VBS order parameter is a complex field, $\Phi_{\rm VBS}$, which is
proportional to the monopole creation operator. Its expectation
value vanishes upon approaching the critical point as $\langle
\Phi_{\rm VBS} \rangle \sim \Delta^{\beta_{\rm VBS}/\nu}$. The
exponent $\beta_{\rm VBS}/\nu$ is the scaling dimension of the
monopole operator, which is expected to be large \cite{gm,senthil1}.
We are interested here in the mixing induced by $\langle \Phi_{\rm
VBS} \rangle$ across the reduced Brillouin zone shown in
Fig~\ref{fig2}. However, because of the shift in the minimum of the
holon-spinon bound state dispersion discussed above, this mixing is
negligible as long as $\sqrt{\delta} < \lambda \sim \Delta^2$, and
so can always be neglected as $\delta \rightarrow 0$. This
establishes the Fermi surface structure claimed in Fig.~\ref{fig2}.

At larger $\delta$, the hole pockets will cross the dashed lines in
Fig.~\ref{fig2}, and Bragg reflections will split the Fermi
surfaces. We analyze this magnitude of the mixing matrix element in
the Hamiltonian,  using the PSG of $\Phi_{\rm VBS}$ in Table 1, in
Appendix~\ref{app:vbs}. We find that it has a value $\sim \Delta
\langle \Phi_{\rm VBS} \rangle \sim \Delta^{1 + \beta_{\rm
VBS}/\nu}$. This splitting of the Fermi surface is negligible
provided the matrix element is smaller than the hole kinetic energy,
or  $ \delta > \Delta^{1 + \beta_{\rm VBS}/\nu}$. Over such a
possible regime of larger $\delta$, the basic pictures of
Figs.~\ref{fig1} and~\ref{fig2} continue to hold. The photoemission
intensity of the ``shadow'' Fermi surfaces noted in Fig.~\ref{fig2}
is proportional to the square of the matrix element or $\sim
\Delta^{2(1 + \beta_{\rm VBS}/\nu)}$

\section{Quantum criticality and conclusions}

To complete the picture, let us address the physics of the shaded
region in Fig.~\ref{fig1}. Here we have a theory of a finite density
of holons $f_{qv}$ interacting with spinons $z_\alpha$ via a U(1)
gauge force $A_\mu$. At sufficiently long scales, the holons will
`Thomas-Fermi' screen that longitudinal $A_\tau$ force, and so
obviate the binding into gauge neutral combination. Consequently the
spinons and holons remain as relatively well-defined excitations,
and we enter a fractionalized holon metal phase. This screening can
be prevented if the spacing between the holons ($\sim
1/\sqrt{\delta}$) is larger than the holon-spin binding length $\sim
1/\Delta$, or $\delta < \Delta^2$; this fixes the boundary of the
shaded region in Fig.~\ref{fig1}. The shaded region will exclude the
unshaded region of the VBS phase where the Fermi surface splitting
is negligible (discussed in the previous paragraph) provided
$\beta_{\rm VBS}> \nu$, an inequality that holds at least for large
$N$.

We now describe the criticality of the hole spectrum at $s=s_c$. We
assume here an observation scale (frequency ($\omega$) or
temperature ($T$)) large enough, or a $\delta$ is small enough, so
that the holes can be considered one at a time. A key observation
about a single hole is that its quadratic dispersion ({\em i.e.\/}
the terms proportional to $1/(2 m_{v\overline{x},\overline{y}})$ in $\mathcal{L}_f$) is an
irrelevant perturbation on the quantum critical point of
$\mathcal{L}_z$ which involves excitations which disperse linearly
with momentum. Consequently, the hole may be considered localized,
and its physics is closely related to the single impurity in a spin
liquid problem analyzed in Ref.~\onlinecite{qimp}. Here we are
interested in the single hole Green's function and this requires the
overlap between states of the spin liquid with and without the
impurity. This can be computed by analyzing the quantum critical
theory of $\mathcal{L}_z$ coupled to one holon localized at $r=0$,
represented by the $r$-independent Grassmanian $f(\tau)$:
\begin{displaymath}
\mathcal{S}_{qc} = \int d^2 r d\tau \mathcal{L}_z + \int d \tau
f^\dagger \left( \frac{\partial}{\partial \tau} + \varepsilon_0 - i
A_\tau (r=0,\tau) \right) f.
\end{displaymath}
Here $\varepsilon_0$ is an arbitrary energy fixing the bottom of the
holon band, and, following earlier arguments \cite{qimp}, the only
relevant coupling between the `impurity' holon degree of freedom and
the bulk degrees of freedom of $\mathcal{L}_z$ is the gauge coupling
associated with $A_\tau$. The spectral function of a physical charge
$e$, $S=1/2$ hole is given by the two-point correlation of the
composite operator $z_\alpha f^\dagger$. If this operator has
scaling dimension $\eta_h/2$, then the universal critical hole
Green's function, $G_h$, is independent of wavevector and obeys the
scaling form
\begin{equation}
G_h = T^{-(1-\eta_h)} \Phi (\hbar (\omega-\varepsilon_0)/k_B T)
\label{gh}
\end{equation}
where $\Phi$ is a universal scaling function. At $T=0$, we obtain an
incoherent spectrum associated with the power-law singularity $G_h
\sim (\omega - \epsilon_0)^{-1+\eta_h}$. We have computed $\eta_h$
by a standard $1/N$ expansion for $G_h$ under the action
$\mathcal{S}_{qc}$ and found
\begin{equation}
\eta_h = 1 - \frac{36}{N\pi^2} + \mathcal{O} \left( \frac{1}{N^2}
\right) \label{etah}
\end{equation}

Another perspective on the exponent $\eta_h$ is obtained by mapping
it to an observable in the lattice non-compact $CP^1$ (NCCP) model
studied by Motrunich and Vishwanath \cite{mv}. On a
three-dimensional cubic lattice with spacetime points $j$, the model
has the complex spinor fields $z_{j\alpha}$ on each lattice site,
and a vector potential $A_{j\mu}$ on each link extending along the
$\mu$ direction from site $j$. The hole Green's function at
imaginary time $M_\tau$ (in units of the lattice spacing) is then
given by the two-point correlator of the $z_{j\alpha}$ in the time
direction along with an intermediate `Wilson line' operator
(representing the contribution of the $f$ fermion) which renders the
correlator gauge invariant (see also Ref.~\onlinecite{wen}):
\begin{displaymath}
G_h (M_\tau) = \left\langle z^{\alpha \ast}_{j} \exp \left( i
\sum_{n=0}^{M_\tau-1} A_{j+n\hat{\tau},\tau} \right) z_{j+M_\tau
\hat{\tau},\alpha} \right\rangle_{\rm NCCP};
\end{displaymath}
this is expected to decay as $e^{- \widetilde{\varepsilon}_0 M_\tau}
/M_\tau^{\eta_h}$ for large $M_\tau$ at the quantum critical point.
The continuum limit of the above correlator was computed by Kleinert
and Schakel \cite{kleinert} in the $1/N$ expansion: their result
agrees with Eq.~(\ref{etah}).

This paper has presented a theory for a strongly interacting quantum
critical point in a metal. The strong interactions imply scaling of
observables as a function of $\hbar\omega/k_B T$, which is often
seen experimentally, but is absent in the weak-coupling theory
\cite{hertz}. The theory has a interesting evolution in the Fermi
surface geometry across the transition, which is reminiscent of that
found in the solution of self-consistent impurity models
\cite{qsi1}. However, the details are different, and can only be
addressed in a quantum field-theoretic framework which we have
described. This work therefore also serves to place studies of
quantum transitions in the `dynamical mean field theory' approach
\cite{qsi1,gabi} in the field-theoretic context.

The $T$ dependent broadening in the quantum-critical hole spectrum
in Eq.~(\ref{gh}) has similarities to recent observations in the
underdoped cuprates \cite{mohit}. The dispersion minimum is shifted
away from the $\vec{K}_p$ points in Fig.~\ref{fig2}, as is the case
experimentally, and in contrast to other theories of the normal
state based upon Dirac fermion spinons. This suggests that a theory
such as ours, based upon a fractionalized Fermi liquid state
\cite{ffl}, should lead to a useful description of the pseudogap
phase.

\acknowledgments

We are grateful to M.~Metlitski for valuable discussions.
This research was supported by the NSF grants DMR-0537077 (SS and RKK),
 DMR-0132874 (RKK) and DMR-0541988 (RKK). AK was
supported by Grant KO2335/1-1 under the Heisenberg Program of
Deutsche Forschungsgemeinschaft, ML by the Harvard Society of
Fellows, and TS by a DAE-SRC Outstanding Investigator Award in
India.


\appendix

\section{Derivation of the PSG}
\label{app:psg}

We are interested here in the physics of the $t$-$J$ like models in
which the Hilbert space is restricted to 3 possible states on each
site of the square lattice: an up/down spin electron, and a vacancy.
The state with two electrons is projected out because of the strong
on-site repulsion, and this constitutes the strong correlation
physics. We can represent these states in terms of canonical
operators by decomposing the electron operator into spinon and holon
operators. On one sublattice of the square lattice we write the
electron operator, $c_\alpha$ as
\begin{equation}
c_\alpha = b_\alpha f^\dagger_+ \label{e1aa}
\end{equation}
where $b_\alpha$ are canonical `Schwinger' bosons and $f_+$ are
canonical fermionic holons. The projection onto the 3 allowed states
is imposed by the local constraint
\begin{equation}
f_+^\dagger f_+ +  b^{\alpha\dagger} b_{ \alpha} = 1 \label{e1}
\end{equation}
on each site. On the other sublattice, we use bosons which transform
as a conjugate respresentation
\begin{equation}
c_\alpha = \varepsilon_{\alpha\beta} \overline{b}^\beta f^\dagger_-
\label{e1a}
\end{equation}
with a similar constraint.

Now consider the mean-field U(1) spin liquid state of the insulator.
This is described by the effective Hamiltonian \cite{aa,rs}
\begin{equation}
H_J =  - Q \sum_{\langle ij\rangle} b_{i\alpha}
\overline{b}_{j}^{\alpha} + \mbox{H.c.} + \lambda \sum_{i}
b_i^{\alpha\dagger} b_{i \alpha} + \lambda \sum_{j}
\overline{b}_{j\alpha}^{\dagger} \overline{b}_{j}^{\alpha},
\end{equation}
where $i$ is restricted to be on one sublattice, and $j$ on the
other, and $Q$, $\lambda$ are positive constants. Any operation
which interchanges the two sublattices leaves $H_J$ invariant under
the PSG mapping
\begin{eqnarray}
b_\alpha &\rightarrow & \varepsilon_{\alpha\beta} \overline{b}^\beta
\nonumber \\
\overline{b}^\alpha &\rightarrow & \varepsilon^{\alpha\beta} b_\beta
. \label{psgb}
\end{eqnarray}
Further, the invariances of Eqs.~(\ref{e1aa},\ref{e1a}) under such
an operation demand that
\begin{eqnarray}
f_+ & \rightarrow & f_- \nonumber \\
f_- & \rightarrow & - f_+. \label{psgf}
\end{eqnarray}
Note especially the sign in the last equation, which is a
consequence of $\varepsilon^{\alpha\beta}\varepsilon_{\beta\gamma} =
-\delta^{\alpha}_{\gamma}$.

The spinon spectrum of $H_J$ shows that the minimum energy
excitations are near zero momentum, and so we may take its low
energy limit by a naive gradient expansion. This leads \cite{rs} to
the effective theory $\mathcal{L}_z$ in Eq.~(\ref{lz}) with
\begin{equation}
z_\alpha \sim b_\alpha + \overline{b}^{\dagger}_\alpha
\end{equation}
The PSG of the $z_\alpha$ in Table~\ref{tablepsg} follows from the
above relations.

Determination of the PSG of the $f_{qv}$ requires the additional
information that the holons have their dispersion minima
$\vec{K}_{1,2}$. The phase factors associated with the
transformation properties of $e^{i \vec{K}_v \cdot \vec{r}}$,
combined with those in Eq.~(\ref{psgf}) lead to the results in
Table~\ref{tablepsg}.

\section{Holes in the N\'eel state}
\label{app:neel}

By spin rotation invariance, we can always rotate the $z_{\alpha}$
condensate (and without {\em any\/} rotation of the spinless
$f_{qv}$) to produce a N\'eel order $N^a = z^{\alpha \ast}
\sigma^{a\beta}_{\alpha} z_\beta$ (where $\sigma^a$ are the Pauli
matrices) polarized along the $(0,0,1)$ direction. Now examine the
response of the theory to a uniform magnetic field $H$ applied along
the $z$ direction. Under such a field, the only change in the action
is that \cite{qimp} $(\partial_\tau - i A_\tau) z_\alpha \rightarrow
(\partial_\tau - i A_\tau +(H/2) \sigma^z ) z_\alpha$. Choosing
$\langle z_\alpha \rangle = \sqrt{|s-s_c|/u} \delta_{\alpha
\uparrow}$, we obtain a term in the Lagrangian
\begin{equation}
\mathcal{L} = \cdots (|s-s_c|/u) \bigl[ iA_\mu - (H/2)
\delta_{\mu\tau} \bigr]^2 + \cdots
\end{equation}
which gaps the $A_\mu$ photon. Integrating out the $A_\mu$ and then
evaluating $M_z = \left. \delta \mathcal{S}/\delta H \right|_{H=0}$,
we obtain an expression for the magnetization density $M_z$
\begin{equation}
M_z = \frac{1}{2} \sum_{qv} q  f^\dagger_{qv} f_{qv} + \ldots
\end{equation}
where the ellipses represent an additional term which measures the
magnetization of the spin waves. This establishes, as claimed in
Fig.~\ref{fig1}, that $S_z = q/2$ for the fermions in the N\'eel
phase.

\section{Holon-spinon bound states}
\label{app:bound}
We describe the structure of a holon-spinon bound state in more
detail. The bound state of a $f_{+1}$ holon with a spinon will be
mixed by $\mathcal{L}_c$ with the bound state of a $f_{-1}$ holon
with a spinon (parallel considerations apply to the $f_{\pm 2}$
holons).

Before writing the wave-function for this bound state, we
need to decompose the relativistic field $z_\alpha$ into
non-relativistic, canonical boson fields $p_\alpha$ (the spinon) and
$h_\alpha$ (the anti-spinon). At low momenta, this is done by the
standard parameterization
\begin{equation}
z_\alpha = \frac{1}{\sqrt{2\Delta}} \left( p_\alpha +
  \varepsilon_{\alpha\beta} h^{\beta\dagger} \right) \label{znr}
\end{equation}

In the absence of $\mathcal{L}_c$ the bound state of $f_{+1}$
and $h_\alpha$ and the bound state of $f_{-1}$ and $p_\alpha$ are
independent, and thus the problem can be solved in the
familiar center-of-mass frame,
\begin{eqnarray}
M_{\overline{j}} &=& m_{v\overline{j}} + \Delta;~~ R_{\overline{j}}  =\frac{m_{v\overline{j}} r_{h\overline{j}} + \Delta r_{s\overline{j}}}{m_{v\overline{j}} + \Delta
} \\
\rho_{\overline{j}} &=& \frac{m_{v\overline{j}} \Delta}{m_{v\overline{j}} + \Delta};~~\vec{r}=  \vec{r}_h - \vec{r}_s
\end{eqnarray}
 $\Delta$ is the mass of the spinon in the non-relativistic
limit, and $m_{v\overline{j}}$ are the holon masses in $\mathcal{L}_f$. The holon
co-ordinate is  represented by $\vec{r}_h$ and that of the spinon by $\vec{r}_s$.
The kets describing the motion of the bound state can be labeled in terms of the momentum $\vec{P}$
of the bound state center-of-mass,
\begin{eqnarray}
|\alpha \vec{P} + \rangle &=& \int d\vec{r_S}d\vec{r_h} e^{i \vec{P} \cdot \vec{R} }
\phi(\vec{r}) f^\dagger_{+1}(\vec{r}_h) h^{\alpha\dagger} (\vec{r}_s) | 0 \rangle\nonumber \\
|\alpha \vec{P} - \rangle &=& \int d\vec{r_S}d\vec{r_h} e^{i \vec{P}
\cdot \vec{R} } \phi(\vec{r}) f^\dagger_{-1} (\vec{r}_h)
p^{\alpha\dagger}(\vec{r}_s) | 0 \rangle \label{ket}
\end{eqnarray}
The kets correspond to the eigenvalues $E^0_{\vec{P}} =
\frac{P^2_{\overline{i}}}{2M_{\overline{i}}}$ and have been written in terms of the
``first-quantized'' ground state wavefunction $\phi(r)$ of the
holon-spinon interaction, $V(r)$. $V(r) = (12 \Delta/N) \ln (r
\Delta)$ binds the holons and spinons over a length scale $\sim
1/\Delta$, with $ \phi (0)\sim \Delta$.

The mixing of the $\pm$ kets due to  $\mathcal{L}_c$ can be studied by interpreting this
term simply as a Hamiltonian $\mathcal{H}_c$ with the operator identification in
Eq.~(\ref{znr}). $\mathcal{H}_c$ only connects kets with the same $\vec{P}$ and $\alpha$
with the  matrix element
$\langle \alpha \vec{P} + | \mathcal{H}_c|\alpha \vec{P} -\rangle = -\frac{\widetilde{\lambda}|\phi(0)|^2P_{\overline{x}}}{M_x}$
(we used the fact that $\partial_x\phi(0)=0$).
Now, by simply diagonalizing a $2\times 2$ matrix we can infer
 the dispersion $E_{\vec{P}}$ induced by $\mathcal{L}_c$ to first
order in perturbation theory in $\widetilde{\lambda}$,
\begin{equation}
E^1_{\vec{P}} =\frac{P_{\overline{i}}^2}{2 M_{\overline{i}}} \pm
\frac{\widetilde{\lambda} P_{\overline{x}} |\phi (0)|^2}{ M_x}
\end{equation}
with eigenmodes which correspond precisely to the electron states in
Eq.~(\ref{psi}) ($x$ is replaced by $y$ for  $f_{\pm 2}$)
. So the bound state dispersion has a minimum at
$K_{\overline{x}}= \pm \lambda$
where $\lambda =  \widetilde{\lambda} |\phi (0)|^2\sim \Delta^2$.
This is responsible for the shift in the centers of
the elliptical hole pockets shown in Fig 2.

\section{Coupling to VBS order}
\label{app:vbs}

Here we describe the terms in action which couple the VBS order
parameter $\Phi_{\rm VBS}$ to the holons and/or spinons. Such terms
are responsible for the mixing between the Fermi surfaces of the VBS
state (shown in Fig.~\ref{fig2}) and the `shadow' Fermi surfaces
obtained by Bragg reflection across the reduced Brillouin zone
boundaries (shown as dashed lines).

The most general terms in the action follow by the requirements that
they be U(1) gauge invariant, and also invariant under the PSG in
Table 1. These requirements turn out to be extremely restrictive.

First, we search for terms which involve $\Phi_{\rm VBS}$ and a
bilinear of the $f_{qv}$ fermions. However a detailed analysis shows
that there {\em no such terms\/} which are invariant under all the
PSG operations; this is seen by first listing the $f_{qv}$ bilinear
invariants under $I_x^{\rm dual}$ (under which $\Phi_{\rm VBS}$ is
invariant), and then noting that their transformations under
$R_{\pi/2}^{\rm dual}$ are incompatible with those of $\Phi_{\rm
VBS}$. Consequently, the coupling between the VBS order and the
fermions will be weaker than might have been initially expected, and
will vanish faster than $\langle \Phi_{\rm VBS} \rangle \sim
\Delta^{\beta_{\rm VBS}/\nu}$ as $\Delta \rightarrow 0$.

The simplest non-vanishing coupling turns out to require the full
$\Psi_p$ electron operators in Eq.~(\ref{psi}). This has the form
\begin{eqnarray}
\mathcal{L}_{\rm VBS} &=& \lambda_{\rm VBS} \Phi_{\rm VBS}^\ast
\Bigl[ -i \bigl( \Psi^\dagger_1 \Psi_4 - \Psi^\dagger_4
\Psi_1  +
\Psi^\dagger_2 \Psi_3 - \Psi^\dagger_3 \Psi_2 \bigr) \nonumber \\
&~&~~~~~~~~~~~+ \bigl( \Psi^\dagger_1 \Psi_2 -
\Psi^\dagger_2 \Psi_1  + \Psi^\dagger_4 \Psi_3 - \Psi^\dagger_3
\Psi_4 \bigr) \Bigr] + \mbox{c.c.}
\end{eqnarray}
Now we need an estimate of the matrix element of $\mathcal{L}_{\rm
VBS}$ between the holon-spinon bound states found in the previous
section. Using Eqs.~(\ref{znr}) and (\ref{ket}), we obtain a
matrix element of order
\begin{equation}
\lambda_{\rm VBS}  \langle \Phi_{\rm VBS} \rangle \frac{|\phi
(0)|^2}{\Delta} \sim \lambda_{\rm VBS} \Delta^{1 + \beta_{\rm
VBS}/\nu}.
\end{equation}
As expected, this does vanish faster than $\Delta^{\beta_{\rm
VBS}/\nu}$, and is responsible for the weak Bragg reflection across
the reduced Brillouin zone boundaries of the VBS state in
Fig.~\ref{fig2}.


%

\end{document}